\documentstyle[12pt, aasms4]{article}
\begin{document}

\vspace*{2.0in}
\title{The state of the molecular gas in a luminous starburst/Seyfert 2 galaxy: NGC 1068 revisited}

\author{Padeli. P. Papadopoulos\altaffilmark{1}}
\affil{Sterrewacht Leiden, P. O. Box 9513, 2300 RA Leiden, The Netherlands}
\and
\author{E. R. Seaquist}
\affil{Department of Astronomy, University of Toronto, 60 St. George st.
       Toronto,\\ ON M5S$-$3H8, Canada.}

\altaffiltext{1}{Department of Astronomy, University of Toronto,
 60 St. George st.  Toronto,\\ ON M5S$-$3H8, Canada.}

\begin{abstract}

 We present fully sampled $ ^{12}$CO,  $ ^{13}$CO J=2--1, 3--2 maps of
 the inner  $\sim 1'\times 1'$ region  of NGC 1068.  We  combine these
 measurements  with an existing   interferometric  map of  $  ^{12}$CO
 J=1--0 that includes single dish data and  thus contains all the flux
 present.    This  allows a  reliable    estimate of  the  $  ^{12}$CO
 (J=3--2)/(J=1--0) ratio at the  highest angular  resolution currently
 possible  and  the use  of  this sensitive   line ratio to  probe the
 physical  condition  of  the   molecular gas.   We  also  present two
 measurements  of the faint  C$  ^{18}$O J=2--1 emission which confirm
 earlier  measurements of  a high $\rm  C  ^{18}O/  ^{13}CO$ intensity
 ratio  in this  galaxy.   The ratios of  the  $  ^{12}$CO, $ ^{13}$CO
 isotopes can only be  reproduced for small/moderate optical depths of
 $  ^{12}$CO J=1--0 ($\tau  \sim 1-2$) which  is incompatible with the
 high   $\rm C ^{18}O/ ^{13}CO$  ratios  observed.  A simple two-phase
 model for the gas can account for all the observed line ratios if the
 C$ ^{18}$O emission  and part of the $  ^{13}$CO emission arise  in a
 dense spatially concentrated component, where  C$ ^{18}$O J=1--0 has
 optical depths  of $\tau \ga 1$.  The  $ ^{12}$CO emission originates
 from  a warmer,  diffuse gas phase  with  $\tau \sim 1-2$ for J=1--0.
 The dense gas phase contains the bulk of the molecular gas mass while
 the diffuse phase may not be virialized leading to an overestimate of
 molecular gas mass when deduced from the luminosity of the $ ^{12}$CO
 J=1--0 line and a standard galactic conversion factor.  This suggests
 that, since type  2  Seyferts harbor a  central starburst  more often
 than type 1, the higher average  $ ^{12}$CO J=1--0 luminosity of type
 2   hinted  by earlier  studies  may  simply reflect a  difference in
 molecular gas excitation rather than in gas mass.

\end{abstract}


\section{Introduction}

 Intense central star  formation   and  possibly an Active    Galactic
 Nucleus (AGN) are  thought to be  the energy sources responsible  for
 the   observed  large IR luminosities  in   bright IRAS galaxies (e.g., 
 \cite{Tel88}; \cite{San91}) where large amounts  of molecular gas are
 found (\cite{San91}).  An  important  problem  in understanding   the
 energetic phenomena occurring in  the nuclear regions of such galaxies
 is determining the  role of  the molecular gas  in  fueling the  AGN
 and/or a   central  starburst.

 Many of  the very IR-luminous  ($\rm L_{\rm FIR}> 10^{11} L_{\odot}$)
 galaxies  are  found to be  strongly  interacting or  merging systems
 (\cite{San88};     \cite{La89}).   Strong   interactions between  two
 galaxies, at least one  of which is  gas  rich, may foster  the rapid
 accumulation of large amounts of molecular gas in the nuclear regions
 (\cite{Ba91})  where they can  ``ignite'' a starburst and ``fuel'' an
 AGN.  For Seyfert galaxies with more  moderate FIR luminosities ($\rm
 L_{\rm  FIR}  \la 10^{11} \  L_{\odot}$)  the picture  is less clear.
 Earlier  work by Heckman et al.   (1989)  suggested that the hosts of
 type 2 Seyfert nuclei  contain, on average,  more molecular  gas mass
 and have  a larger  FIR   luminosity and  hence higher   star forming
 activity than type 1.  More recent studies by  Maiolino et al. (1997)
 of a larger and better defined sample  of Seyferts and field spirals
 confirms the higher  FIR luminosities of  Seyfert 2's  but casts some
 doubts  on the  purported  difference in molecular  gas mass  between
 these galaxies and Seyfert 1's, or  field spirals.  In the same study
 preliminary evidence  points towards bars, interactions and distorted
 morphologies as being present more frequently in  the hosts of type 2
 than in  type  1 Seyferts.  In  this  case, efficient  molecular  gas
 transport  towards the nucleus will  be present more  often in former
 rather than the latter.   Hence, provided that this mechanism remains
 effective over scales ranging from  $\rm L\sim 1$  kpc down to a  few
 parsecs, it  can  naturally  explain why   the  onset of   a  central
 starburst correlates with a higher probability of obscuring the Broad
 Line Region of the AGN.

Nevertheless  such  a scenario  still cannot   explain a larger  total
molecular    gas reservoir in   Seyfert 2  with  respect  to Seyfert 1
galaxies.  However, since type  2 are more likely  to harbor a central
($\sim  1-5$  kpc) starburst than  Seyfert  1's (Maiolino 1995), there
exists the possibility that any difference in their average $ ^{12}$CO
J=1--0 luminosities is caused by differences in the average excitation
of the molecular gas rather than the total gas mass.

 NGC 1068 is   the  closest and best  studied  example  of a  luminous
 Seyfert 2  galaxy  which  also  harbors an intense   starburst in its
 central  $\sim  $2.5 kpc.  It  has a  large  bolometric luminosity of
 $\sim 3\times 10^{11}$ L$_{\odot }$  roughly equally divided  between
 the unresolved    AGN and the  starburst disk    (\cite{Tel84}).  The
 distribution  of  molecular  gas    is characterized  by  a   nuclear
 concentration within the central 100 pc  (Planesas, Scoville \& Myers
 1991;  Helfer \& Blitz 1995) where  mostly dense gas ($\rm n(H_2)\sim
 10^{5}\ cm^{-3}$) resides (Jackson et al.  1993; Tacconi et al. 1994;
 Helfer \& Blitz 1995) and  a more extended molecular gas distribution
 that coincides  with the starburst region   and consists of  two inner
 spiral arms and a bar (Helfer \& Blitz 1995).  The nuclear component
 is thought  to   contribute  to the   built-up  of an   AGN-obscuring
 molecular torus  (Planesas et al.  1991; Cameron et al. 1994; Tacconi
 et al.   1994).   In  the present  work  we  use  this  galaxy   as a
 ``laboratory'' to study  the excitation of  the bulk of the molecular
 gas in the  intense starburst  environment of the   host galaxy of  a
 Seyfert 2 nucleus.   To do  so  we use  our  fully sampled maps of  $
 ^{12}$CO, $   ^{13}$CO J=3--2, 2--1  acquired   with the James  Clerk
 Maxwell  Telescope (JCMT)\footnote{The JCMT  is operated by the Joint
 Astronomy Center   in  Hilo,   Hawaii   on   behalf  of  the   parent
 organizations   PPARC in the    United Kingdom, the National Research
 Council of Canada and the The Netherlands Organization for Scientific
 Research.}    in  combination  with   an  existing interferometric  $
 ^{12}$CO J=1--0 map (\cite{Hel95}).

 Resolution-matched  observations  of different rotational lines  of $
^{12}$CO and  especially $ ^{13}$CO  provide an excellent probe of the
physical conditions  in the bulk of  the molecular  gas.  A ratio like
(3--2)/(1--0)  can be    particularly  sensitive   to gas   excitation
conditions since the J=3 level lies 32 K above ground versus 5.5 K for
J=1 while the J=3--2   transition has a significantly higher  critical
density  ($\rm n_{\rm cr}=4 \times  10^4$  cm$ ^{-3}$) than the J=1--0
transition ($\rm n_{\rm cr}=2\times 10^3$ cm$ ^{-3}$) in the optically
thin regime.  Moreover the  maps of the rare  $ ^{13}$CO isotope allow
the study of  the molecular gas using  less opaque lines than the ones
of $ ^{12}$CO and  therefore sensitive to  the average conditions of a
larger fraction of molecular material.

Throughout this study  we assume  a distance  to  NGC 1068  of D=14  Mpc
 (Sandage \& Tammann  1981, $\rm H_{\circ }   = 75$ km  s$ ^{-1}$) for
 which 1$''$ corresponds to 68 pc.

\section{OBSERVATIONS}

 We used the receivers A2, B3i  and the newly commissioned receiver B3
on JCMT  in several observing runs  in order to  produce fully sampled
maps  (Nyquist sampling) of the emission  lines $ ^{12}$CO, $ ^{13}$CO
J=3--2, 2--1 of the  inner $\sim 1'\times 1'$  region of NGC 1068.  We
also observed the very weak C$ ^{18}$O J=2--1 transition in two points,
the details of all our observations are summarized in Table 1.
\vspace*{1.0cm}

\centerline{EDITOR: PLACE TABLE 1 HERE}
\vspace*{0.3cm}

The  map center  was located at  $\alpha (\rm  B1950)  = 02^{\rm h}  \
 40^{\rm m} \ 07^{\rm s}.06$, $\delta(\rm B1950)= -00^{\circ } \ 13' \
 31.45''$ and the receiver was tuned to a velocity of $\rm v_{\rm LSR}
 =1125 $ km s$ ^{-1}$.  For more efficient mapping of the CO emission,
 the area  mapped was along a  rotated grid with $\rm PA=40^{\circ }$.
 The grid-cell   is an equilateral triangle   with  a size  of $\Delta
 \theta  _{\rm  s} \sim  \theta   _{\rm HPBW}/2$, where  $\theta _{\rm
 HPBW}$  is the HPBW  of the  gaussian  beam.  This grid provides  the
 optimum coverage of the CO-emitting region of NGC~1068 at a Nyquist
 sampling and is  uniquely determined by  the  demand of a)~a~uniform
 sampling pattern  and,  b)  every sampled point   having  the maximum
 number (N=6)   of the closest neighboring   points  at a  distance of
 $\Delta \theta _{\rm s}$.  We performed  all our observations using 
 beam  switching with  a  beam  throw  of  150$''$ in  azimuth at  the
 recommended rate of 1 Hz and only for the very weak C$ ^{18}$O J=2--1
 the  rate  was  2 Hz in  order  to  achieve  a stable  baseline.  The
 pointing  and  focus  were monitored  frequently  by observing bright
 quasars, planets and OMC1 (for $ ^{13}$CO J=3--2).

We converted the $\rm  T^{*} _{\rm A}$ temperature  scale of  the JCMT
 spectra to the   $\rm T_{\rm mb}$   scale by using the relation  $\rm
 T_{\rm mb}=T^{*}_{\rm A}/\eta _{\rm  mb}$.  We  performed measurements of
 $\eta _{\rm mb}$ for all the receivers  by observing Saturn (A2, B3i)
 and  Jupiter (B3). We found  excellent  agreement (within $\sim 10\%$)
 with  the values  reported  in the JCMT   manual (Matthews 1996). The
 adopted beam efficiencies are  $\eta  _{\rm mb}(\rm A2)=0.69$,  $\eta
 _{\rm mb}(\rm B3i)  = 0.55$ and $\eta _{\rm  mb}(\rm  B3)=0.62$.  The
 error  associated with the line  intensity measurements was estimated
 using the relation

\begin{equation}
  \left(\frac{\delta   \rm T}{\rm T}\right)=\left[  \left(\frac{\delta
        \rm  T}{\rm  T}\right)^2 _{\rm  ther}+ \left(\frac{\delta  \rm
        T}{\rm T}\right)^2  _{\rm   calib} +  \left(\frac{\delta   \rm
        T}{\rm T}\right)^2 _{\rm syst} \right]^{1/2}.
\end{equation}
 
The first  two terms express the thermal  rms error and the stochastic
 uncertainty of the calibration of  the  line intensities.  The  third
 term  is  the   systematic   uncertainty of the   assumed   telescope
 efficiency  factors.   If T is the   main beam brightness temperature
 averaged over $\rm N_{\rm ch}$ channels and $\rm  N_{\rm bas}$ is the
 total number of   channels defining a  symmetric baseline  around the
 spectral line, then

\begin{equation}
  \left(\frac{\delta \rm T}{\rm  T}\right)_{\rm  ther} =  \frac{\delta
    \rm    T_{\rm ch}}{\rm T} \left(\frac{\rm     N_{\rm bas} + N_{\rm
        ch}}{\rm N_{\rm bas} N_{\rm ch}}\right)^{1/2},
\end{equation}

\noindent
where $\delta  \rm  T_{\rm  ch}$ is the   thermal noise  per  channel.

The calibration  uncertainty factor was  estimated from the dispersion
of the observed intensities of many  spectral line standards with high
S/N,  and it  was found to  be  $\sim 0.10-0.15$ for  all the observed
frequencies.   The  third factor  is $\sim  0.10$  for all frequencies
(Friberg,  Sandell, private  communication).   Finally we performed an
additional check on the overall uncertainty of the line intensities by
frequently monitoring the $ ^{12}$CO lines  in some standard points of
our map.  For the weak  $ ^{13}$CO lines we  monitored the spectrum of
OMC1 at $\rm V_{\rm LSR}=1125$ km s$ ^{-1}$ where various bright lines
are still  present.  In all cases  we found excellent agreement of the
line profiles and a measured dispersion of line intensities consistent
with the one estimated from the first two terms of Equation 1.

\section{DATA REDUCTION}

  The JCMT  data were reduced  using the SPECX  reduction package.  We
 smoothed all our spectra  to a common  velocity resolution of $\Delta
 \rm V_{\rm chan}\sim 8$ km s$ ^{-1}$ and removed linear baselines. In
 the case of the $ ^{12}$CO J=3--2 transition we also removed some bad
 channels that were present at the high velocity end of the band.

The grid  maps were interpolated  by convolving  them  with a gaussian
function.  For the J=3--2 maps the  FWHM of the interpolating function
used was $\theta _{\rm  i}=8''$.   The interpolated map  has effective
resolution     of     $[\theta _{\rm   i}^2     +   \theta   ^2  _{\rm
HPBW}(3-2)]^{1/2}=16''$.   For  the J=2--1 grid  maps  we used $\theta
_{\rm i}=11''$ and the effective resolution of the interpolated map is
$[\theta   _{\rm i}^2 +  \theta ^2  _{\rm HPBW}(2-1)]^{1/2}\sim 24''$.
The maps were  further analyzed with  the AIPS reduction package.  The
task HGEOM was employed to re-grid the JCMT maps in order to re-orient
them in   an  (RA, DEC)  grid.   A  check  of  the consistency of  the
interpolation  and   re-gridding  was   performed   by comparing   the
area-averaged $\rm T_{\rm mb}$ in  individual channel maps between the
original grid maps and the final ones  used in our analysis.  We found
good agreement within the thermal rms uncertainties.   The maps of the
velocity-averaged main beam brightness $\langle \rm T_{\rm mb} \rangle
_{\Delta \rm v}$ for $ ^{12}$CO, $ ^{13}$CO are shown in Figures 1 and
2 for the J=2--1 and 3--2 transitions respectively.

\placefigure{fig1}

\placefigure{fig2}

The $ ^{12}$CO J=1--0 map used in  our analysis was obtained by Helfer
 \&  Blitz (1995)   with  the  Berkeley-Illinois-Maryland  Association
 (BIMA) interferometer  and the NRAO 12m  telescope.  This  map is the
 most  suitable  one  for comparison with   our  JCMT  maps since  its
 combination of interferometer  and single  dish measurements recovers
 all  the CO  flux.  The interferometer   map has  been corrected  for
 primary  beam attenuation   by   assuming  a  gaussian  profile  with
 HPBW=100$''$.  We  convolved the $ ^{12}$CO  J=1--0 channel maps to a
 spatial resolution  of 16$''$ and  converted the  original brightness
 units  (mJy/beam) to the $\rm T_{\rm  mb}$ temperature scale in order
 to compare  them to the $ ^{12}$CO  J=3--2 map.  The channel maps and
 the  associated  line ratio  $\rm   r_{32}=(3-2)/(1-0)$ are shown  in
 Figure 3.
\vspace*{-0.25cm}

\placefigure{fig3} 
  
The close correspondence of  the $ ^{12}$CO J=3--2  and the $ ^{12}$CO
J=1--0 emission in every channel demonstrates that no serious pointing
offsets    are   present between     the  two   maps.   The  estimated
channel-to-channel rms dispersion of the $\rm  r_{32}$ ratio is of the
order of $10\%$.  This includes only  the thermal rms  error since the
other two sources  of error for   the JCMT maps  (Equation  1) and the
systematic  calibration uncertainty in  the flux  conversion factor of
the BIMA $ ^{12}$CO J=1--0 map are constant across  the passband.  The
largest single source  of error  is  the $\sim 30\%$ flux  calibration
uncertainty of the  $  ^{12}$CO J=1--0 BIMA  map.   This results to  a
total uncertainty of $\rm 35\%$ for the $\rm r_{32}$ ratio.

The  C$ ^{18}$O J=2--1 spectra correspond  to  a spatial resolution of
$\theta _{\rm HPBW}=22''$.  In order to compare them to the $ ^{13}$CO
spectra towards the same locations we  used an interpolated $ ^{13}$CO
J=2--1  map with  $\theta _{\rm  G}=8''$,  which   gives an  effective
resolution of   $\theta _{\rm i}=23''$.   The  spectra, smoothed  to a
common frequency resolution of 25  MHz (34 km  s$ ^{-1}$) are shown in
Figure 4.
\vspace*{-0.25cm}

\placefigure{fig4}

\section{RESULTS}

The $   ^{13}$CO, $  ^{12}$CO  J=2--1 and  3--2  maps  (Figures  1, 2)
demonstrate that the relative intensities of these two isotopes change
significantly  across the  emitting  region.  There  is  a significant
increase of the intensity of $ ^{13}$CO relative to $ ^{12}$CO towards
the location   of the giant  molecular gas  associations (GMAs) in the
southern part of  the central  starburst region of  NGC 1068.    These
molecular  gas concentrations   show    up in  high  resolution   maps
(Planesas,  Scoville \&  Myers 1991;  Helfer  \& Blitz  1995)  as high
brightness regions with   large  inferred  H$_2$  surface   densities.
Especially large variations of  $ ^{12}$CO/$ ^{13}$CO are observed for
the J=1--0 transition (Helfer \& Blitz 1995; Papadopoulos, Seaquist \&
Scoville 1996).  Large  variations of this  ratio have   been recently
reported by Aalto et al. (1997) in the extreme starburst/merger system
Arp 299 where  high  resolution interferometric measurements  indicate
changes by a factor of $\sim 6$.

The  $\rm r_{32}$ ratio  also  shows a  large  range of  values in the
various regions in the inner 3~kpc of NGC 1068  (Fig.  3).  The lowest
values $\rm r_{32}\sim 0.3-0.4$ occur towards the eastern part of the
emission while the highest ones  $\rm r_{32}\sim 0.6-0.8$, towards the
western and  southwestern part  where  the bright  GMAs are located.
Moreover, the ratio  $\rm r_{21}$ varies  in a similar fashion with  a
range of $\rm  r_{21} \sim 0.6-0.7$  and $\rm  r_{21}\sim 0.7-0.8$ for
the eastern   and   the western/southwestern parts   of   the emission
respectively.

 Line ratios as low as $\rm r_{32}\sim  0.3-0.4$ characterize the bulk
of the relatively cold gas in the Giant Molecular Clouds (GMCs) in the
disk of  the  Galaxy  (\cite{San93}).  The  same   is true   for  $\rm
r_{21}\la  0.7$, usually measured in  the   outer disk  of the  Galaxy
(\cite{Has97}), or the  disks  of other galaxies (e.g.,  \cite{Eck90};
Eckart et al.~1991).  An LTE approximation of such low ratios yields a
$\rm T_{\rm kin}\la   6$ K, comparable to  the  minimum  value of  7~K
permitted by cosmic  ray heating (Goldsmith \&  Langer  1978).  Then a
maximum brightness  of $ ^{12}$CO J=1--0  would be $\rm T_{\rm b} \sim
3$ K, much smaller what is usually measured in NGC 1068 (\cite{Pl91}),
which implies  sub-thermal excitation of  the J=3--2, 2--1 transitions
towards the regions with such low $\rm r_{32}$ and $\rm r_{21}$ ratios.

  On  the other hand ratios  as   high as  $\rm r_{32}\sim 0.6-1$  are
  measured  towards  central   regions  of   starbursts (\cite{Dev94};
  \cite{Wil92}) and close to the cores of star forming GMCs like W3 in
  the Milky  Way (e.g., \cite{Phil81}).   A  ratio of $\rm  r_{21}\sim
  0.80-1.0$  is found  mainly towards the   centers of nearby  spirals
  (\cite{BrCo92}),     centers  of   starbursts (e.g.,   \cite{Eck90};
  \cite{Eck91}; \cite{Wil92}), close  to HII regions  and star forming
  cores in the Galaxy (\cite{Has97}).   In NGC 1068 the largest values
  ($\rm r_{32}=0.8$ and  $\rm  r_{21}=1$)  are observed  towards   the
  location of the  most massive GMAs  identified in the interferometer
  maps.   Under  LTE conditions these    ratios imply average  kinetic
  temperatures of $\rm T_{\rm kin}  \ga 25$~K.  In contrast, the  bulk
  of the molecular gas in the plane  of the Milky Way is significantly
  colder at $\rm  T_{\rm kin}\approx  10$ K (\cite{San93}),  undoubtly
  the result   of  a  more  quiescent  environment   than  the central
  starburst region of NGC~1068.

\subsection{The isotope line ratios}

As reported previously  the    value of $ ^{12}$CO/$   ^{13}$CO  ratio
changes across the central kpc-size  starburst region of NGC 1068 with
the largest variations measured for the J=1--0 transition.  This shows
most prominently in high  resolution interferometer maps (Papadopoulos
et al. 1996) but even   in the area-averaged  spectra shown in Figure
5.

\placefigure{fig5}

 These spectra demonstrate that the  ratio $\rm R_{10} =$$\rm ^{12}CO/
^{13}CO$ (J=1--0) changes  by a  factor of $\sim   2$  along the  line
profile  while variations  of $\rm  R_{21}$  (J=2--1) and $\rm R_{32}$
(J=3--2)   follow in  a similar  fashion   but are significantly  less
prominent.  It is important to note that the variation of $\rm R_{10}$
across the  line  profile   is  not  influenced by     the  systematic
uncertainties  of the line intensities,  since they can only introduce
an overall shift  of the $\rm T_{\rm mb}$  scale.  The line shapes are
subject only to the thermal rms error across  the bandpass which is of
the order of  $\sim 10\%$  for the J=1--0,   and $\sim 15\%$  for  the
J=2--1 and 3--2 transitions.

The reason for using the OVRO $ ^{13}$CO J=1--0 spectrum (Papadopoulos
et al. 1996), which does not include single  dish data, instead of the
one from BIMA/NRAO 12m (Helfer \& Blitz 1995) stems  from the fact the
later has  a significantly lower S/N, a  result of poor weather during
the  observations  (Helfer, private communication).  Nevertheless both
data sets  exhibit   the same basic characteristics.    Moreover,  the
largest variation of $\rm R_{10}$ is found  among the low and the high
velocity end of  the J=1--0 line  profile, where  the addition  of the
single flux data is  not expected to  alter the line intensity (Helfer
\& Blitz~1995).

In the inner kiloparsec-size  region of NGC  1068 we find $\rm R_{J+1\
 J}  \ga  10$  (J=0, 1,~2)  throughout, irrespective   of location or
 averaging  scale.  For the entire range   of the $\left[ \rm ^{12}CO/
 ^{13}CO\right]$   abundance values (e.g.,   Langer  \& Penzias  1993;
 Henkel \& Mauersberger 1993) such high  isotopic ratios correspond to
 low $ ^{12}$CO optical depths.  Assuming LTE and $\left[ \rm ^{12}CO/
 ^{13}CO\right]=40$, a value of $\rm R_{10}\ga 10$  corresponds to a $
 ^{13}$CO J=1--0  optical depth of  $\tau^{(13)}  _{10}\la 0.1$, and
 hence rather  high  $\rm r^{(13)}  _{21}=r_{21}(  ^{13}CO)$ and $\rm
 r^{(13)} _{32}=r_{32}( ^{13}CO)$ ratios  (or equivalently, high  $\rm
 R_{21}$, $\rm  R_{32}$  ratios) with  respect to the ones  observed.
 This  has  been  used to support    the notion  of two-phase   gas in
 starburst centers (Aalto  et  al.  1995),  yet there is  still a wide
 range of {\it non-LTE} conditions that must be explored before such a
 conclusion can  be reached.  We will see  that  the best evidence for
 such a state of the  molecular gas comes  from the C$ ^{18}$O  J=1--0
 (Papadopoulos et al.  1996) and J=2--1 data (this work).

\subsection{The high C$ ^{\bf 18}$O/$ ^{\bf 13}$CO ratio}

The ratio $\rm R^{(18)} _{21} =  C ^{18}O/ ^{13}CO$ J=2--1 obtained at
the two positions   is   $\rm R^{(18)}  _{21}=0.26\pm0.07$   and  $\rm
R^{(18)} _{21}=0.20\pm0.06$ (Figure  4).  These values, while somewhat
lower,    are  comparable to      the   average ratio  $\rm   R^{(18)}
_{10}=0.3\pm0.1$ found from high resolution OVRO maps (Papadopoulos et
al.  1996) for the J=1--0 transition.   Moreover, these maps also show
that the C$ ^{18}$O emission does not smoothly trace $ ^{13}$CO in the
various GMAs and varies drastically from one location to the next with
values  as    high  as  $\rm   R^{(18)}  _{10}\sim   0.65$.    This is
significantly  higher  than  the    abundance ratio $\rm    [C ^{18}O/
^{13}CO]=0.12-0.14$ inferred for   the Milky Way  from various studies
(e.g.,  \cite{Wan80};   \cite{Henk93}; \cite{Lan93}).     Ratios  $\rm
R^{(18)} _{10}\ga 0.2$ are also found in several central starbursts of
galaxies (e.g., \cite{Sag91}; \cite{Cas92}; \cite{Henk93}).

While it  is  possible that   an  enhancement of  the  $\rm  [C^{18}O/
 ^{13}CO]$  abundance can  occur in starburst  environments (Henkel \&
 Mauersberger 1993) and hence play a role in  the observed high values
 of $\rm R^{(18)} _{10}$ and $\rm R^{(18)} _{21}$, it is very unlikely
 that it can be  solely responsible for the  high $\rm R^{(18)} _{10}$
 values  observed in    individual GMAs (Papadopoulos   et  al. 1996).
 Furthermore, the large spatial variations of  the $\rm R^{(18)} _{10}$
 ratio over scales of $\sim  400$~pc are difficult to attribute solely
 to a varying $\rm [C^{18}O/ ^{13}CO]$ abundance because the mixing of
 interstellar gas is expected to be effective enough to homogenize the
 isotope ratios over similar or  larger scales, even during an ongoing
 burst of star formation (Henkel \& Mauersberger 1993).

Thus, it seems more likely that the optical depth of the rare isotopes
is not negligible. For $\rm R^{(18)} _{10}\ga 0.20$, the optical depth
of  $ ^{13}$CO J=1--0   is  $\tau^{(13)} _{10}\ga 1.5$  (LTE assumed),
which directly contradicts the general conclusion reached earlier from
the $\rm R_{10}$ ratio which yields $\tau^{(13)} _{10}\la 0.1$.  This
constitutes the strongest evidence that $ ^{12}$CO,  $ ^{13}$CO and C$
^{18}$O cannot be all tracing the same phase of the molecular gas.

\section{DISCUSSION}

Line ratios of CO  and its isotopes as  inputs to a radiative transfer
model that solves the rate equations for the various rotational levels
are useful probes  of the  physical  conditions of the molecular  gas.
Obviously the larger the  number  of the input  ratios the  better the
constraints on the gas properties.  The various spectral lines for NGC
1068 have been observed at very different angular resolutions with the
lowest one ($\theta _{\rm HPBW}\sim 24''$ for J=2--1) being comparable
to the source size.   Hence, using the spatially-averaged spectra over
the entire CO  emitting area ($\sim  1'\times 1'$) allows the estimate
of all the  line  ratios for that  area  and thus offers  the  maximum
number of inputs that can be used to constrain the physical properties
of the molecular gas.  These line ratios are tabulated in Table 2.
\vspace*{1.0cm}

\centerline{EDITOR: PLACE TABLE 2 HERE}
\vspace*{0.3cm}

For  the modeling  of  the molecular  gas  properties we  used a Large
Velocity  Gradient (LVG) code.  We  searched an  extensive grid of LVG
models  with a    parameter  range of  $\rm T_{\rm    kin}  =10-120$ K
($\Delta\rm T_{\rm kin}=2$~K), $\rm n(H_2)=(0.1-10^4)\times  10^3$~cm$
^{-3}$ ($\rm \Delta  logn(H_2)=0.5$) and,  $\Lambda \rm  =  X/(dV/dr)=
(0.1-10^2)\times  10^{-6}$ (km s$  ^{-1}$~pc$ ^{-1}$)$ ^{-1}$ ($\Delta
\rm log\Lambda = 0.5$), where $\rm X  = \left[ ^{12}CO/H_2\right]$ and
$\rm  dV/dr$ is the  velocity gradient  for the ``average''  molecular
cloud.  We assumed an abundance ratio  of $\rm [ ^{12}CO/ ^{13}CO]=40$
since studies of the  Milky Way (Langer \&  Penzias 1993)  and central
regions of starburst   galaxies   (Henkel at   al.  1993; Henkel    \&
Mauersberger 1993) show this value  to  be a  suitable average of  the
abundances observed towards  galactic centers.  The abundance  of $\rm
[C^ {18}O/ ^{13}CO]=0.15$ is derived from $\rm [ ^{12}CO/ ^{13}CO]=40$
and  $\rm [  ^{12}CO/C ^{18}O]=250$, which  is  the highest such value
derived for molecular clouds in the Milky Way  (Wannier 1980), and was
measured towards its  center.  We examine the  entire grid of  the LVG
models for the values of $\rm r_{21}$, $\rm r_{32}$ that correspond to
$0\%$, $-30\%$  and   $+30\%$ offset in   the BIMA  $  ^{12}$CO J=1--0
intensity.  The best fit is  found for the  highest $\rm r_{21}$, $\rm
r_{32}$ ratios  which correspond to the $30\%$  reduction  in the flux
density of the $ ^{12}$CO J=1--0 map.

Apart   from giving the  best fit,  such an  offset  of the $ ^{12}$CO
J=1--0  intensity observed by BIMA  seem to be  corroborated by single
dish  measurements reported by Maiolino  et  al.  (1997) and Kaneko et
al.      (1989).  The   first   used    the NRAO   12m   to  obtain  a
velocity-integrated main  beam  brightness  for the  inner $55''\times
55''$ of  NGC 1068 which is  $\rm I=(90\pm 13)K  \ km \ s^{-1}$, while
the latter used the NRO 45m telescope to map  a similar area and found
$\rm I=(83\pm  17) K\  km\ s^{-1}$.  These   values were obtained from
their data after converting  the $\rm T^{*} _{\rm  R}$ (NRAO  12m) and
$\rm T^{*} _{\rm A}$ (NRO 45m) temperature scales to $\rm T_{\rm mb}$.
The integrated  flux found over the same  region from the  BIMA map is
$\rm I=(131\pm  40)  K\ km  s^{-1}$.  Clearly  only a large  offset of
$\approx -30\%$ can bring the  BIMA value in  agreement with the other
two.  Moreover, adopting  such  an  offset yields $\rm   r_{21}=0.97$,
which  in  much  better   agreement with  the   value $\rm r_{21}=1.1$
reported by Braine \& Combes (1992) over the same angular size.

The conditions corresponding to the best fit are $\rm T_{\rm kin}=20$
 K, $\rm n(H_2)=10^4$ cm$ ^{-3}$ and $\Lambda = 10^{-6}$ (km s$ ^{-1}$
 pc$ ^{-1}$)$  ^{-1}$.  It is interesting to  note that Thronson et al
 (1987)  using    mm/sub-mm  continuum  measurements deduce   a   dust
 temperature of T=20 K  for the inner 55$''$ of  this galaxy, which is
 similar to the gas  temperatures we find from  the LVG fits.  This is
 expected for well mixed gas and dust where the $ ^{12}$CO, $ ^{13}$CO
 lines trace the same ``average'' conditions as the mm/sub-mm emission
 originating from dust.

As expected from  our previous discussion, the  best LVG  model cannot
reproduce the observed high  C$ ^{18}$O/$ ^{13}$CO line  ratios, since
the deduced optical depth of the $  ^{12}$CO J=1--0 transition is $\rm
\tau _{10} \sim 1-2$.   In this case $ ^{13}$CO,  C$ ^{18}$O  are very
optically thin and the  ratio of their intensities  is equal  to their
assumed relative abundance.  The   moderate  optical depth of the    $
^{12}$CO J=1--0 transition  was suggested by  Aalto et  al.  (1995) as
being  one of the  characteristics of a  second, more diffuse and warm
gas phase  that surrounds a  dense and more spatially concentrated one
in the nuclear regions  of starbursts (see  also Wall et al.  1993 for
earlier  work on spirals).  This would  naturally explain why both the
average brightness of the $ ^{12}$CO  J=1--0 transition as well as the
$\rm R_{10}$ line ratio are more  sensitive functions of velocity (and
therefore of  location  within the   galaxy)  than  the  J=2--1,  3--2
transitions (Figure 5)  and   the $\rm  R_{21}$, $\rm  R_{32}$  ratios
respectively.

For $\rm T_{\rm    kin}\geq 20$ K  it  is  $2.7\tau  _{10}  \leq  \tau
 _{21}\leq 4\tau _{10}$  and $3\tau _{10}  \leq \tau _{32}\leq 9  \tau
 _{10}$ (LTE), hence  a  moderate $\tau_{10}\sim 1-2$  corresponds  to
 $\tau_{\rm J+1\ J}\sim 3-6$  (J=1,2) or higher.   This range of $\tau
 _{10}$  values  allows  the   $  ^{12}$CO  J=1--0   transition to  be
 significantly more sensitive than the J=2--1 and J=3--2 to variations
 of  gas column   density and  excitation  conditions.  Higher  values
 $\tau_{10}(>2)$  would render  all  three transitions optically thick
 and  thermalized even at     moderate/low ($\rm n(H_2)\la  10^3$  cm$
 ^{-3}$) gas  densities because of  radiative trapping, and hence they
 would be equally ``insensitive''  to the conditions of the  molecular
 gas except the kinetic temperature.  On the other hand, significantly
 lower  values  $\tau  _{10}(\la   0.1)$ yield  small/moderate optical
 depths  for all  three transitions  and not just  J=1--0.  Hence they
 will all   be sensitive to  the gas  excitation conditions  and their
 intensities  and $\rm R_{\rm J+1\  J}$ line ratios will vary strongly
 along the line profile, reflecting the changing excitation conditions
 in the inner region of NGC 1068.

An important  consequence of the  moderate  $ ^{12}$CO  J=1--0 and the
correspondingly  small ($< 0.1$) $ ^{13}$CO   J=1--0 optical depths is
that   the latter transition can  be  slightly super-thermally excited
(Goldsmith 1972; de Jong  et al.  1975; Leung  \& Liszt 1976).  Indeed
for the best LVG solution it is $\rm  T_{\rm ex}(1-0)>T_{\rm kin}$ for
$  ^{13}$CO.   In this  case the  sensitivity  of $\rm  R_{10}$ to the
excitation conditions   with respect  to   the $\rm  R_{21}$  and $\rm
R_{32}$ ratios becomes more pronounced while $\rm r_{21}\sim~0.8-1.0$,
and $\rm r^{(13)} _{21}\la 1.3$ can still be  fitted with a single gas
phase of   small/moderate   $ ^{12}$CO  J=1--0  optical   depth  ($\rm
R_{10}\ga  10$).  Later  we  will  see that  this property   of the  $
^{13}$CO J=1--0 transition remains  as one the characteristics  of the
diffuse gas phase in a two-component gas model.

\subsection{The heterogeneous gas in NGC 1068}

The inability of a single gas phase to  account for the line ratios of
 the $ ^{12}$CO, $  ^{13}$CO and C$ ^{18}$O  isotopes in the starburst
 region of NGC 1068 suggests at least two  distinct phases.  One phase
 dominating the $ ^{12}$CO while the  other is responsible for most of
 the C$ ^{18}$O emission, and with the $ ^{13}$CO having contributions
 from both.

The notion  of a two-component  or generally multi-component molecular
gas phase seems trivial  in the face of numerous  studies of the Milky
Way and other galaxies that reveal  a large range of temperatures $\rm
T_{\rm kin}(\sim 10-100 K)$  and densities $\rm n(H_2)(\sim  10^2-10^8
~cm ^{-3})$  for  the   molecular gas.    This  notion is   indeed not
particularly useful  unless  one includes the  spatial scale involved.
Examination of the physical conditions  of the molecular gas on scales
of $\rm L\la   1-50$ pc is  bound to  yield a very   different picture
depending  on whether  one looks at  Orion-type  molecular clouds that
harbor ongoing star formation or more  quiescent ones.  Averaging over
much larger ($\rm  L\ga 0.5$ kpc)  scales  ``smoothes'' out  the local
density and temperature irregularities to a degree that a narrow range
of    temperatures   and  densities,  deduced     from molecular  line
spectroscopy, can fit the observed line ratios.

{\it It is when such ``averaging'' fails to converge towards a single
narrow set of physical conditions that the notion of more than one
molecular gas components becomes meaningful and necessary.}

  Numerous studies of molecular gas in other  galaxies (e.g., Knapp et
al.  1980; Wall \&  Jaffe 1990; Eckart  et al.  1991; Braine \& Combes
1992; Aalto et al.  1995) have demonstrated a broad differentiation of
molecular gas  properties with warm  ($\rm  T_{\rm kin}\ga 20$  K) and
dense gas ($\rm n(H_2)\ga 10^4$~cm$  ^{-3}$) lying in the inner  $\sim
0.5-1$ kpc,   while colder ($\rm  T_{\rm kin}  \la 10$ K),  less dense
($\rm  n(H_2)\la 10^3$~cm$ ^{-3}$) gas  is located  further out in the
disk.  A similar differentiation seems to be in place in the Milky Way
(e.g., Bally et al.  1988;  Stark et al.   1989,  1991; Binney et  al.
1991; Spergel \& Blitz 1992).  The properties  of the molecular clouds
in galactic  disks seem to  be easily  described  by a single  set  of
molecular gas properties  or  equivalently the ``averaging''  of those
properties over scales of $\rm L\ga  0.5$ kpc converges to single type
of ``average'' cloud.

This does not seem  to be the case for  the molecular gas lying in the
 central galactic regions.  Several studies (e.g., Wall \& Jaffe 1990;
 Eckart  et al. 1990;  Wall et al. 1993;  Aalto  et al. 1995) indicate
 that even when averaged  over $\sim 0.5-1$  kpc, the line emission of
 the various trace molecules does not point towards a single gas phase
 but at least two.  For starburst  centers this ``differentiation'' of
 the molecular gas seems to become more acute (Aalto et al.  1995) and
 is thought to be driven by a variety of mechanisms, namely an intense
 UV field, large   turbulent linewidths, tidal forces and  cloud-cloud
 collisions. In this picture one  phase consists of warm and  diffuse
 and low-density     molecular gas originating   from  the  heated and
 disrupted outer  envelopes of molecular   clouds while the  other one
 consists of the more ``protected'' interiors of such clouds where the
 gas is denser and colder.

\subsection{A two-phase model for NGC 1068}

 The wealth of molecular line data gathered  for the central starburst
region  of  NGC~1068 allows   for  another  rigorous  test  of   the
aforementioned  ideas and a better determination  of the properties of
the two gas phases.

We denote as (B) a dense, concentrated gas  phase where $ ^{13}$CO and
even  C$ ^{18}$O may have  substantial  optical depths   and as (A)  a
diffuse phase where $\tau _{10}\sim 1-2$  for $ ^{12}$CO J=1--0.  Then
all the observed line ratios  can be expressed as  a weighted mean  of
the  ratios of these two phases.   Hence the $\rm R^{(18)} _{10}$ line
ratio can be expressed as follows:

\begin{equation}
\rm  R^{(18)}  _{10} = \frac{\rm   R^{(18)}  _{10}(A)+[f_c \rho _{13}]
R^{(18)} _{10} (B)} {\rm 1+[f_c \rho _{13}]},
\end{equation}

\noindent
where $\rm f_c$ is  the filling factor  of phase (B) relative to phase
(A) and $\rho _{13}=\rm T_{\rm R}(B)/T_{\rm R}(A)$ is the ratio of the
$ ^{13}$CO J=1--0 brightness temperature for the two gas phases.

In phase (A) both  $ ^{13}$CO  and  C$ ^{18}$O isotopes  are optically
thin (so there is little radiative trapping), and  since they have the
same collisional excitation  coefficients  and are  influenced by  the
same background  radiation field it  is $\rm R^{(18)} _{10}(A)=\left[C
^{18}O/   ^{13}CO\right]$.  Assuming    the   highest such  abundance,
measured   towards the  central   regions   of the  Milky~Way   (e.g.,
\cite{Wan80};   \cite{Henk93};   \cite{Lan93}),  means $\rm   R^{(18)}
_{10}(A)=0.15$. Hence, $\rm R_{10}$ can be expressed as

\begin{equation}
\rm R_{10}=\frac{\rm R_{10}(A)+[f_c \rho _{13}] R_{10}(B)}{\rm 1+ [f_c \rho _{13}]}.
\end{equation}

The rest of  the line ratios $\rm  r_{\rm  J+1\ J}$ and $\rm  r^{(13)}
_{\rm J+1\ J}$ (J=1, 2) for $ ^{12}$CO and $ ^{13}$CO respectively can
be expressed as follows:

\begin{equation}
\rm r_{\rm J+1\ J}=\frac{\rm r_{\rm J+1\ J}(A)+ [f_c \rho _{12}] r_{\rm J+1\ J}(B)}{\rm 1+[f_c \rho _{12}]}
\end{equation}

\begin{equation}
\rm r^{(13)} _{\rm J+1\ J}=\frac{\rm r^{(13)} _{\rm J+1\ J}(A)+ [f_c \rho _{13}] r^{(13)} _{\rm J+1\ J}(B)}{\rm 1+[f_c \rho _{13}]},
\end{equation}

\noindent
where  $\rm   \rho  _{12}  = T_{\rm  R}(B)/T_{\rm   R}(A)$ denotes the
brightness temperature  ratio of  $ ^{12}$CO  J=1--0  for the two  gas
phases. The two factors $\rho  _{13}$ and $\rho  _{12}$ are related by
the simple relation

\begin{equation}
\rho _{12}=\rm \left[ \frac{R_{10}(B)}{R_{10}(A)}\right] \rho _{13}.
\end{equation}

We find that the  range of the  observed line ratios can be reproduced
only when $\rm  f_{\rm  c} \rho  _{13}\la 1$  and $\rm f_{\rm  c} \rho
_{12}\ll 1$.  The latter yields $\rm r_{21}\approx r_{21}(A)$ and $\rm
r_{32}\approx   r_{32}(A)$  (Equation   5),  i.e.,  these   ratios are
dominated by  the  diffuse phase (A).   The C$  ^{18}$O J=1--0 optical
depth  for phase (B)  is found  to be $\tau  ^{(18)} _{10}\sim  1$ and
hence $  ^{13}$CO  and $ ^{12}$CO  in this  phase are optically thick.
Radiative trapping is then enhanced and can  thermalize the $\rm J\leq
3$, transitions of $ ^{12}$CO at densities of $\rm n(H_2)\sim 10^2\ cm
^{-3}$, and $  ^{13}$CO at  $\rm  n(H_2) \sim 10^3\ cm^{-3}$.  non-LTE
effects are expected to be  important for the  $ ^{13}$CO, C$  ^{18}$O
transitions arising in the diffuse phase~(A).

In this  simple model  we assume LTE  excitation of  the $ ^{12}$CO, $
^{13}$CO J=1--0, 2--1, 3--2 and  C$ ^{18}$O J=1--0 transitions for the
more  spatially concentrated  gas phase   (B),  while making no   such
assumption for the ``envelope'', diffuse  gas phase (A).  The  deduced
line ratios for  the latter are analyzed using  LVG modeling  with the
same   parameter   space and abundance  ratios     as in our  previous
single-phase LVG analysis.  A good solution for phase (A) is found
only for  $\rm  r_{21}=0.97$, $\rm  r_{32}=0.74$,  corresponding to  a
$-30\%$  offset in the BIMA  $ ^{12}$CO J=1--0 intensity.  The results
for both gas phases are summarized in Table 3.

\vspace*{0.5cm}

\centerline{EDITOR: PLACE TABLE 3 HERE}
\vspace*{-0.5cm}

\subsection{The diffuse gas phase}

From Table  3   we  see that  this   gas  phase is  characterized   by
small/moderate optical $ ^{12}$CO J=1--0  depths of $\tau _{10}\la 1$,
relatively low density $\rm n(H_2)\sim 10^3$ cm$ ^{-3}$ and probably 
higher  kinetic temperatures  than  the  denser  and  optically thick
phase (B).  The   $ ^{13}$CO J=1--0   transition is now  significantly
super-thermal, which is responsible   for  the high $\rm R_{10}=   17$
 and moderate $\rm r_{21}\sim 1$ and $\rm r^{(13)} _{21}\sim 1.4$
ratios.  This non-LTE  effect can occur  for a wide range  of physical
conditions as long  as $\rm   T_{\rm  kin}\ga 20$ K, $\rm    n(H_2)\ga
2\times 10^3$ cm$ ^{-3}$   and  $\tau^{(13)} _{10}\la 0.1$  (Leung  \&
Liszt 1976). Moreover, the higher the $\rm T_{\rm kin}$ the larger the
density  regime over which   superthermality  occurs, e.g.,  for  $\rm
T_{\rm kin}=60$ K  and  $\Lambda   \sim 10^{-5}$  (km s$  ^{-1}$~pc$
^{-1}$)$ ^{-1}$ the  density range is  $\rm  n(H_2)\sim 10^3-10^5$ cm$
^{-3}$.  This includes densities well  past the thermalization density
of the J=1--0 transition ($\sim 2\times 10^3$ cm$ ^{-3}$).

These conditions are more likely to be found in starburst environments
 rather than quiescent  ones.  There the  intense UV radiation and the
 turbulent motions present will  heat and disrupt the molecular clouds
 and thus create the  warm diffuse gas  phase where $ ^{13}$CO  J=1--0
 may  become  super-thermally   excited.  The  line   ratios are  then
 sensitive to $\rm (n(H_2), \Lambda )$ but they generally satisfy $\rm
 r_{21}\sim   0.8-1.1$,  $\rm r^{(13)}   _{21}\sim  1.2-2.2$  and $\rm
 R_{10}\ga 10$, which contains the range of the values found for these
 ratios by Aalto et al. (1995) for starbursts.

  The high value of  the $\rm R_{32}$ ratio   for phase (A) (Table  3)
significantly restricts the LVG  parameter space.  This stems from the
fact that the  conditions  for  superthermality  are bounded  by   the
conditions needed to  keep a low $\tau  _{32}$ (and hence a  high $\rm
R_{32}$) which call for a low excitation of the  J=3 level, i.e., $\rm
n(H_2)< 4\times  10^4$~cm$   ^{-3}$ and/or  $\rm T_{\rm  kin}<  30$ K.
However,  in   a starburst  environment  a    wide  range of   kinetic
temperatures  $\rm T_{\rm kin}(\ga 30)$ K  can be expected, especially
for the  ``exposed'' molecular  gas in the   outer areas  of molecular
clouds  (Aalto et al.  1995). Hence the aforementioned restrictions to
the excitation  conditions of the    diffuse gas phase  become  mainly
restrictions on its density, namely $\rm 10^{3}\ cm^{-3}\la n(H_2) \la
10^4\ cm^{-3}$.

\subsubsection{Gas kinematics and the range of dV/dr}

The other important characteristic of the diffuse gas phase is that is
 probably  not virialized.  This can be  viewed as  a direct result of
 the various mechanisms that create this phase by disrupting initially
 denser and more compact molecular clouds.  After the ``envelope'' gas
 phase is created the energetic  environment  in which it resides  may
 not allow    it  to settle   to   self-gravitating  structures.   Its
 prevailing kinematic state is important since the velocity-integrated
 luminosity of $ ^{12}$CO J=1--0 is routinely  used as an indicator of
 gas  mass in galaxies with one  of the key assumptions (among others)
 being that   molecular  gas consists of   an   ensemble of virialized
 clouds, hence linewidth  reflects  mass.  For   $  ^{12}$CO emission
 dominated  by   non-virialized   diffuse  gas    this  will   clearly
 overestimate the gas mass.

 The LVG approximation  is valid when the  linewidth $\Delta \rm v$ of
the average cloud is larger than the  thermal one, i.e., $\rm \Delta v
\gg \Delta  v _{\rm th}$.  For  temperatures of $\rm  T_{\rm kin} \sim
10-50$ K the thermal linewidth  is $\rm \Delta v_{\rm th} =0.1-0.2$~km
s$  ^{-1}$.  Hence  LVG approximation   is expected to   be valid  for
linewidths of  the order of $\rm  \Delta v \ga 10\times  \Delta v_{\rm
th} =   1-2$ km\ s$  ^{-1}$,  which  is what  is  usually  observed in
individual molecular clouds.  An upper limit  to the linewidth of such
clouds  is set by the fact  that their individual linewidths cannot be
larger than the observed  ones in extragalactic systems, which consist
of many molecular clouds.   For  a typical extragalactic linewidth  of
$\rm \Delta v_{\rm G} = 300$~km s$ ^{-1}$ consisting of the linewidths
of at least $\sim  10$ clouds,  the  maximum linewidth is $\rm  \Delta
v_{\rm max} = 0.1 \times \Delta v_{\rm G} = 30$ km s$ ^{-1}$.  This is
a  rather generous upper limit  since the  velocity dispersion seen in
large individual Galactic GMCs is of the order  of $\Delta \rm v= 3-8\
km\ s^{-1}$ (Scoville \& Good 1989).   For a cloud  size of $\rm L\sim
0.5-50$ pc,  the extreme velocity range  of  $\sim 2-30$ km  s$ ^{-1}$
yields a maximum range of $\rm dV/dr \sim  (0.04-60)$~km~s$ ^{-1}$~pc$
^{-1}$ for the average velocity gradient per cloud.

  If the ``average'' cloud is  virialized then, for  a given average gas
density, the expected range  for $\rm dV/dr$ is significantly narrower.
For a spherical cloud of radius R the virial theorem gives:

\begin{equation}
\rm \left(\frac{\rm dV}{\rm dr}\right)_{\rm VIR} \approx \frac{\delta v _{\rm vir}}{
\rm 2R}=\left(\alpha \frac{\pi G \mu}{3}\right)^{1/2}\ \langle n \rangle ^{1/2},
\end{equation}

\noindent
where $\mu $ is the mean mass per particle  and $\langle n \rangle$ is
the mean number density of the cloud. The factor $\alpha \sim 0.5-2.5$
depends    primarily    on   the  assumed      density profile    (see
Bryant~\&~Scoville~1996).  Expressing the last result in astrophysical
units gives

\begin{equation}
\rm \left(\frac{\rm dV}{\rm dr}\right)_{\rm VIR} \approx 0.65\alpha ^{1/2}\
 \left(\frac{\rm \langle n \rangle}{10^3\rm cm^{-3}}\right)^{1/2}\
km \ s^{-1}\ pc^{-1}.
\end{equation}

For NGC  1068, the best  LVG solution  for  phase (A)  gives (Table~3)
$\langle  n \rangle =  3\times 10^3$~cm$ ^{-3}$,  so  for the largest
reasonable  value   of     $\alpha    =  2.5$   we    estimate    $\rm
\left(dV/dr\right)_{\rm VIR}\sim 2$ km s$  ^{-1}$ pc$ ^{-1}$. The same
model also gives $\Lambda =3\times 10^{-6}$ (km s$ ^{-1}$ pc$ ^{-1}$)$
^{-1}$.  For $\rm \left[  ^{12}CO/H_2\right]=10^{-4}$ this yields $\rm
dV/dr = 33$ km s$  ^{-1}$ pc$ ^{-1}$, which  is $\sim 17$ times larger
than    the    virial    value.     Hence,   unless    the  $\left[\rm
^{12}CO/H_2\right]$ abundance  is significantly  lower in the  diffuse
phase, such a large value  of dV/dr  suggests  that this phase in  not
virialized.  A similar situation is found  for the starbursting center
of IC~342 by Irwin \& Avery 1992,  which they interpret  in terms of a
significantly lower $\rm  \left [ ^{12}CO/H_2\right]$ abundance rather
than   in terms of   non-virialized  gas.   In Ultraluminous  Infrared
Galaxies, such a gas  phase seems to be  filling the entire intercloud
medium (Solomon et al.  1997) in the inner few hundred parsecs.

In  all  these cases the    $  ^{12}$CO  linewidth   will have   other
 significant contributions besides the  gravitational potential of the
 gas itself, namely pressure, stellar and other CO-dark mass (Downes et
 al.   1993)  and  the general potential  of  the  galaxy (Solomon  et
 al. 1997).

\subsection{The dense gas phase}

A lower limit for  the gas density of phase  (B) can be deduced  from
 the information in  Table 3 and the  fact that LTE  provides a good
 description of the excitation of the  gas in this phase.  The optical
 depth of the C$ ^{18}$O J=1--0 transition can be expressed as follows

\begin{equation}
\tau  ^{(18)}  _{10}=1.23\times   10^4    \left(\frac{\rm     1-e^{\rm
-5.26/T_{\rm   kin}}}{\rm      T_{\rm  kin}}\right)  \left[\frac{\rm C
^{18}O}{\rm H_2}\right]      \rm    n(H_2)  \left(\frac{\rm    dV}{\rm
dr}\right)^{-1},
\end{equation}

\noindent
where $\rm T_{\rm kin}$ is the kinetic temperature of the gas in (B) phase.

Assuming that the ``average'' cloud of this gas  phase is a virialized
structure means that  $\rm dV/dr$ is  given by Equation  9, so  for an
abundance $\left[\rm ^{12}CO/H_2\right]=10^{-4}$ we get

\begin{equation}
\rm n(H_2)= 2.78\times 10^{-4} \alpha \left(\frac{\rm T_{\rm kin}}{\rm
1-e^{-5.2 6/\rm T_{\rm kin}}}\right)^2\ \left[\frac{\rm ^{12}CO}{\rm C
^{18}O}\right]^2 (\tau^{(18)} _{10})^2.
\end{equation}

A temperature  of  $\rm T_{\rm  kin}= 10$ K  and  an  optical depth of
$\tau^{(18)} _{10}=0.7$  are  the minimum values characterizing  phase
(B).   These   values  and   an  abundance    of  $\left[\rm ^{12}CO/C
^{18}O\right]=\left[\rm    ^{12}CO/   ^{13}CO\right]\times   \left[\rm
^{13}CO/C   ^{18}O\right]=40\times   1/0.15\approx 267$   yield a  gas
density of $\rm n(H_2)\approx 10^4$~cm$ ^{-3}$ ($\alpha = 1.5$).  This
is  a minimum value since $\rm  T_{\rm kin}$  can attain significantly
larger  values  (Table 3)   which would yield   larger values  of $\rm
n(H_2)$, e.g., for $\rm  T_{\rm kin}=20$ K and $\tau^{(18)} _{10}=1.1$
it is $\rm n(H_2)\approx 3\times 10^5$~cm$ ^{-3}$.  Such densities are
comparable to  the average densities found in  the  cores of molecular
clouds  like Orion, M  17 and Cepheus by  Bergin  et al.  (1996) which
also    find good agreement  between the   virial  masses and the ones
deduced  from   C$ ^{18}$O  observations  (Goldsmith  et  al.   1997).
Situations where even  the rare C$  ^{18}$O isotope  may have moderate
optical depth and is associated  with  high density molecular gas  are
known.   In these cases  C$  ^{18}$O shows  good kinematic and spatial
correspondence with much higher dipole  moment molecules like  HC$_3$N
(Bergin  et   al. 1996)   and  NH$_3$  (Fuente~1993).   Similarly high
densities but with higher average temperatures  are found also towards
the Galactic Center region (Guesten 1989).

The large abundance ratio $\left[ \rm ^{12}CO/C ^{18}O\right]$ plays a
 crucial   role in  making  these  two isotopes  suitable  for broadly
 tracing the two gas phases described previously.  Indeed, while it is
 frequently said  that molecules with high   dipole moments (e.g., CS,
 HCN) are  needed in order to   trace high density  gas,  to a certain
 extent, rare CO  isotopes  can  do  the same.   This  happens because
 unlike  the   high  density tracing  molecules,  $   ^{12}$CO has  an
 abundance high enough for moderate/high optical depth to arise in the
 general conditions of the ISM.  Most of the  emission of a transition
 arises  in  regions where   $\tau   \ga 1$   and  the  transition  is
 thermalized.   Thus $ ^{12}$CO J=1--0  can  trace a diffuse gas phase
 since, for $\tau _{10} \sim 1-10$, radiative trapping will thermalize
 this transition at $\rm n_{\rm A}(H_2)=n_{\rm crit}\times (1-e^{-\tau
 _{10}})/\tau  _{10}\approx  10^2-10^3\  cm^{-3}$,   where $\rm n_{\rm
 crit}$  is the critical density   of  the J=1--0 transition.  On  the
 other  hand,   assuming uniform  gas excitation,   the  less abundant
 isotope C$  ^{18}$O with $\rm \left[ ^{12}CO/C ^{18}O\right]=250-500$
 (Wannier 1980), will  attain an optical depth  of $\tau _{10} \sim 1$
 in regions with densities of  $\rm n_{\rm B}(H_2)\sim (250-500)\times
 n_{\rm A}(H_2)\ga 2\times 10^4\ cm^{-3}$.

\subsection{The changing physical conditions of the gas}

The different  angular  resolution of  the  maps   of the  various  CO
transitions prevent  us from  using  all the observed line   ratios to
study the molecular gas within  the starburst region  of NGC 1068 at a
common, high angular   resolution.   Nevertheless, the wide range   of
physical conditions  can be revealed  from the  $\rm r_{32}$ and  $\rm
R_{10}$ ratios alone and  simple  one-phase LVG modeling.   Apart from
the  fact that both these ratios  are sensitive to gas excitation they
can  be  estimated at  the  highest possible  resolutions,   and hence
because of less  spatial smoothing they  can be sensitive to  a wider
range  of physical conditions.   For  $\rm R_{10}\ga 10$ ($\langle \rm
R_{10} \rangle=14$)  the  range   of $\rm  r_{32}$    ($\sim 0.4-0.8$)
corresponds to  $\rm T_{\rm kin}\geq  20$K, $\rm n(H_2)=10^3-10^4$~cm$
^{-3}$  and $\Lambda=10^{-6}-10^{-5}$    (km s$ ^{-1}$   pc$  ^{-1}$)$
^{-1}$.  For  $\rm  r_{32}=0.57-1.1$ (--30\%   offset for $   ^{12}$CO
J=1--0) this parameter space becomes even wider.  In all cases we find
the most dense gas inferred for the large GMAs  in the western part of
the central emission.

Using  all  the global line    ratios and our   simple two-phase model
confirms   the  results  of  the  one-phase   LVG  analysis.   For the
``low-velocity''  CO emission  ($\rm  V_{\rm LSR}\sim  980-1050$~km~s$
^{-1}$) corresponding  to GMAs in the easternmost  part of the central
region (Helfer \& Blitz 1995)  we find negligible contribution of  the
dense gas phase (B) to the $ ^{12}$CO and the $ ^{13}$CO emission, and
the diffuse phase  (A) seems to  be  the dominant component.   For the
``high-velocity'' CO  emission ($\rm V_{\rm  LSR}\sim 1200-1280$ km s$
^{-1}$) originating from the   GMAs  located in the westernmost   part
there is a significant contribution from the dense and optically thick
gas phase (B).   This picture is  further supported by a sensitive map
of  the HCN   J=1--0  which traces   dense  gas  made  with   the IRAM
interferometer  (Tacconi et   al.   1994)  at  a  resolution of  $\sim
4''\times 3''$.  This  map shows that, apart  from the Seyfert nucleus
itself, the brightest HCN emission originates from the western part of
the inner kpc-size region of  NGC 1068 and  is most prominent  towards
the  location  of    the  massive GMA   at    $(\Delta \alpha,  \Delta
\delta)\approx (-8'', -15'')$ from the center.

 Our  study lacks sufficient angular resolution  to shed more light to
the physical conditions of the nuclear  gas ($\rm L\la 100$ pc), where
warm ($\rm T_{\rm kin}\ga 70$ K) and dense  ($\rm n(H_2)\sim 10^5$ cm$
^{-3}$)  gas dominates  (Tacconi   et  al.   1994).  As we   discussed
earlier, even the  $\rm r_{32}$ ratio at  $\sim 16''$, is dominated by
the extended CO emission from the GMAs throughout.

\subsection{Molecular gas mass in NGC 1068}

The   most   common method  used    to  find molecular   gas  mass  in
 extragalactic  systems relies  on  the   so-called standard  galactic
 conversion factor  $\rm X_{\rm G}$  which converts  the luminosity of
 the $ ^{12}$CO J=1--0 line to gas mass, namely

\begin{equation}
\rm X_G = \frac{M(H_2)}{L_{\rm CO}}=5.0\ M_{\odot }\ (km\ s^{-1}\ pc^2)^{-1},
\end{equation}

\noindent
with its numerical value adopted from Solomon \& Barett (1991).  The $
^{12}$CO J=1--0 luminosity $\rm L_{\rm CO}$ is estimated from

\begin{equation}
\rm L_{\rm CO}=A _{\rm s}\  \int _{\Delta \rm v} \int _{\rm A_{\rm s}}
 T _{\rm b}(a, v)
\ da dv = \Omega _{\rm c} \ D^2 \int _{\Delta v} T^{(\rm a)} 
_{\rm b} (v) dv,
\end{equation}

\noindent
where $\rm T^{(\rm a)}  _{\rm b} (v)$   is the area-averaged  spectrum
(see Figure 5) and D the distance to the galaxy.

The  velocity-integrated  brightness  of  $  ^{12}$CO J=1--0 from  the
Kaneko et al.  (1989),  Maiolino et  al.   (1997) and our  measurement
(scaled by $-30\%$) give an average of $\rm \langle I \rangle = 88\ K\
km \ s^{-1}$, over  an area of $\sim 1'\times  1'$.   Using $\rm  X_G$ then
gives $\rm M(H_2)\approx 6\times 10^9\ M_{\odot}$.

Under certain assumptions and taking  into account only differences in
 average  density $\langle \rm n(H_2)  \rangle $ and $ ^{12}$CO J=1--0
 brightness temperature $\langle \rm T_{\rm b}\rangle $ (see Bryant \&
 Scoville 1996 for a more  complete treatment), the conversion  factor
 $\rm X$ can be expressed as follows (e.g Radford et al. 1991)

\begin{equation}
\rm X = 2.1\ \frac{\sqrt{\langle \rm n(H_2) \rangle}}{\rm \langle T_b \rangle }.
\end{equation}

\noindent
  From our single-phase LVG  analysis we find  that $\rm X/X_G \approx
1-3$, which can yield gas mass estimates as high as $\rm M(H_2)\approx
1.8\times 10^{10}\ M_{\odot }$.

Several studies (Dickman et al. 1986; Maloney \& Black 1988; Bryant \&
Scoville 1996) attempt  to quantify the  effects of the molecular  gas
environment on the value of X, especially when the $ ^{12}$CO emission
is optically  thick.  These  studies indicate  that, when  a  standard
galactic conversion  factor is used, the  presence of a non-virialized
gas  component will  result  to an overestimate    of gas mass,  while
significant shadowing  of molecular clouds  in spatial and/or velocity
space to an underestimate.  Except when the angular resolution is high
enough  to resolve   individual clouds, cloud-cloud  shadowing may  be
important  only towards very  quiescent cloud  environments unlike the
ones  expected  in  a  central  starburst.   Therefore  the sole  most
important factor affecting  X is the  presence of a non-virialized gas
component where molecular  cloud  linewidths are  no longer good  mass
indicators.

A different method is to use rare isotopes of CO like $ ^{13}$CO or C$
^{18}$O to  deduce molecular  gas mass.   The main  assumption here is
that these isotopes have small/moderate  optical depths throughout the
volume of the emitting gas.  Then the total number  of the CO  isotope
molecules at the rotational level J is given by the expression

\begin{equation}
\rm N_J = \frac{8\pi k \nu _{\rm J, J-1} ^2}{\rm h c^3 A_{J, J-1}}
\ \beta _{J, J-1} ^{-1}\ L_{\rm CO}(J, J-1),
\end{equation}

\noindent
where $\rm A_{J,  J-1}$ and  $\beta _{\rm  J, J-1}$  are the  Einstein
coefficient  and the escape  probability of the $\rm J\rightarrow J-1$
transition (e.g., \cite{sob60}; Castor   1970) respectively, and  $\rm
L_{\rm CO}(J, J-1)$ is the corresponding luminosity  (Equation 13).  A
good approximation of the total $\rm H_2$ mass is then provided by

\begin{equation}
\rm M(H_2)=X_{\rm CO}\left(N_1+N_2+N_3\right) \mu m_{\rm H_2},
\end{equation}

\noindent
where $\rm X_{\rm  CO}$ is the abundance of  $\rm H_2$ relative to the
particular CO isotope and $\mu  \rm m_{\rm H_2}$  is the mean mass per
$\rm  H_2$  molecule. Hence the expression   for  the $\rm X_{\rm IS}$
factor that converts  the   J=1--0 luminosity  of the  CO  isotope  to
molecular gas mass is as follows

\begin{eqnarray}
\hspace*{-0.5cm} \rm X_{\rm IS}  =   \frac{\rm 8\pi k \mu m_{\rm H_2}}{\rm h c^3}
\left(\frac{\nu^{2} _{10}}{\rm A_{10}}\right) \left[1+\left(\frac{\nu _{21}}{\nu _{10}}\right)^2
\frac{\rm A_{10}}{\rm A_{21}} \ \frac{\beta _{10}}{\beta _{21}} r_{21}  + 
\left(\frac{\nu _{32}}{\nu _{10}}\right)^2\frac{\rm A_{10}}{\rm A_{32}}\
 \frac{\beta _{10}}{\beta _{32}} r_{32}\right] \beta^{-1} _{10}  \rm X_{\rm CO}, 
\end{eqnarray}

\noindent
where  $\rm   r_{J+1\ J}=L_{\rm CO}(J+1,   J)/L_{\rm CO}(1,0)$. 
After substitution this equation yields

\begin{equation}
\rm   X_{\rm IS}  =    0.078   \left[r\right]     \left[1+0.42\left(\frac{\beta
_{10}}{\beta _{21}}\right)r_{21} +  0.26\left(\frac{\beta _{10}}{\beta
_{32}}\right)r_{32}\right] \beta^{-1} _{10},
\end{equation}

\noindent
where  we  assumed $\rm  \left[H_2/  ^{12}CO\right]=10^4$ and $\rm \left[r
\right]=\left[ ^{12}CO/ ^{\rm x}C ^{\rm y}O\right]$ is the abundance of
$ ^{12}$CO relative to the isotope used.

The assumption about the virialization of  the average molecular cloud
needed to deduce X (Equation 16) is no longer necessary.  The omission
of  $\rm N_J$ for  J=0 and $\rm J>3$  is not expected  to have a major
effect on   the  estimated $\rm  M(H_2)$   for the average  conditions
encountered  in molecular  clouds.  This  is suggested by   a study of
Giant Molecular Cloud  (GMC) cores done  by Goldsmith et al.   (1997),
where  they find good  agreement between the virial  mass and the mass
deduced using this method and observations of  the C$ ^{18}$O isotope.
They also find that the true mass can be at most $\sim 2$ times larger
than the one  computed from Equation 16.  This  occurs for  cold ($\rm
T_{\rm kin}\approx 10$ K) and  diffuse ($\rm n(H_2)<10^3$~cm$  ^{-3}$)
gas where  the term $\rm N_0$ becomes  important, or warm ($\rm T_{\rm
kin}>50$ K) and dense ($\rm n(H_2)>10^5$~cm$  ^{-3}$) gas where levels
above J=3 become significantly populated.

In the   optically thin case,  $\beta  _{\rm J, J-1}\sim  1$, and $\rm
 X_{\rm  IS}$  is independent  of the details    of the gas excitation
 conditions  and velocity field  that  determine the  value of  $\beta
 _{\rm J+1, J}$'s.  Of course in the optically thick case these become
 important and more careful analysis  is needed to estimate the escape
 probabilities.

Applying this method for $ ^{12}$CO or $ ^{13}$CO transitions, we find
a gas mass  of $\rm M(H_2)\approx 5\times 10^8\  M_{\odot }$ from both
isotopes.  This  value  is  $\sim  10-35$ times  smaller than  the one
deduced  from the standard galactic conversion  factor $\rm X_{\rm G}$
or its density/temperature  ``corrected''    value X.  Such    a large
discrepancy could not be due only to the fact that Equation 16 gives a
lower limit to  the molecular gas  mass.  Summing up the population of
all the energy levels  can result in a gas  mass that is at most $\sim
2$ times  larger  (Goldsmith et al.   1997),  which still  leaves $\rm
M(H_2)$ at  least $\sim 5$ times smaller  than the  one found from the
standard galactic conversion factor.

Two  effects that can  contribute to this  difference in the estimated
gas mass are:  a) the presence of  a non-virialized gas component and,
b) a large optical depth of the particular CO isotope used.  The first
effect will lead to an overestimate of the molecular gas mass whenever
a factor that converts $ ^{12}$CO line luminosity to gas mass is used.
Our  analysis  suggests   that such  a   non-virialized  molecular gas
component  indeed exists and dominates  the $ ^{12}$CO emission in the
central region of NGC 1068.

The second effect is due to the fact that in an optically thick medium
significant amounts of gas mass  will contribute only a small fraction
of the total   observed  luminosity.    This is  straightforward    to
understand when the line formation mechanism is local because then the
observed emission comes from the outer  layer of the cloud where $\tau
\sim  1-3$ and a lot of  gas mass can  ``hide''  in its interior.  The
$\beta _{\rm J+1, J}$'s factors can, to  a certain extent, correct for
this effect    but  obviously this  correction  becomes   increasingly
unreliable the larger the optical depth.

 However  in the case  of large  scale  systematic motions within each
cloud (and among the various clouds themselves), the various optically
thick components appear   centered at different velocities   along the
line  profile  and   hence  are ``visible''.    In   such  a situation
significant gas  mass can still    ``hide''  provided that the    more
optically thick component has a lower  temperature, a smaller velocity
gradient and overall volume than the more diffuse gas component.  This
is exactly the type of molecular gas differentiation found for the gas
in NGC 1068  and generally in  starburst centers.  A combination  of a
small spatial/velocity filling factor and  a lower average temperature
can reduce  the luminosity  contribution of  the  optically thick  gas
phase to $\la   0.1\%$ of the  luminosity of  the  diffuse, warmer gas
component for equal amounts of gas mass.

The  more accurate  description of  the  molecular gas  as a two-phase
system allows a better estimate of the total gas mass as well as an
estimate of the relative gas mass contained in each phase, namely 

\begin{equation}
\frac{\rm M_B}{\rm M_A}=0.13\left(\rm f_c \rho _{13}\right)
\left[\frac{\rm ^{13}CO}{\rm C ^{18}O}\right] \frac{\left(\rm T_{\rm kin}\
e^{5.26/\rm T_{\rm kin}}\right)\beta^{-1} _{10}}{\rm 1+0.42\ r^{(13)} _{21}
+0.26\ r^{(13)} _{32}}\ \rm R^{(18)} _{10}(\rm B),
\end{equation}

\noindent
where $\rm T_{\rm kin}$ and    $\beta _{10}$ are the average   kinetic
temperature and  escape probability of C$  ^{18}$O J=1--0 in gas phase
(B)  and $\rm r^{(13)}  _{21}$ and $\rm  r^{(13)}  _{32}$ are the line
ratios for the optically thin $ ^{13}$CO ($\beta _{\rm J+1\ J}\sim 1$, 
J=0, 1, 2) in phase (A).

Typical values  from  Table 3  yield  $\rm M_{\rm B}/M_{\rm  A}\approx
5-10$ suggesting that most of the mass resides in the dense, optically
thick   gas    phase (B).   The estimated   total    gas mass  is $\rm
M(H_2)=M_{\rm  A}+M_{\rm  B}\approx    3\times 10^9\   M_{\odot}$.  As
expected, this estimate yields  a mass that  is larger (by a factor of
$\sim 6$)  than the  one found  from Equation 18 under the assumption
that $ ^{13}$CO is optically  thin throughout the emitting volume, yet
smaller (by a factor  of $\sim 2-6$)  than the ones estimated from the
standard galactic conversion factor.

\subsubsection{Estimates of $\rm M(H_2)$ in starburst environments}

Our study of  the IR-luminous starburst in  NGC 1068  suggests that in
the   inner   few   kiloparsecs, the  highly   excited, non-virialized
molecular gas component that dominates  the $ ^{12}$CO emission can be
the  cause of a systematic  overestimate of molecular  gas mass when a
standard galactic  conversion   factor is  used.   In  cases  of other
IR-luminous galaxies with powerful central starbursts like Mrk 231 and
NGC  7469 the gas mass deduced  from the  standard galactic conversion
factor is manifestly  overestimated since  it  yields a mass  that  is
comparable or larger than the dynamical mass contained within the same
volume (Bryant \& Scoville 1996; Genzel et al.  1995).  While this can
still  be the result of a  particular geometry  of the CO-emitting gas
(e.g., for Mrk 231,  Bryant \& Scoville  1996) it is also likely  that
the diffuse  gas component in their starburst  regions is in  a highly
excited state with high  temperatures and large non-virial linewidths.
In this case the velocity-integrated $ ^{12}$CO J=1--0 luminosity will
be significantly larger than  in more quiescent  environments, without
this reflecting a proportionally larger molecular gas mass.

In  the particular case  of Seyferts, since type  2 are more likely to
harbor starbursts than type 1 or field spirals (Maiolino et al. 1995),
it follows   that molecular gas   mass deduced  from their  $ ^{12}$CO
J=1--0 luminosity will    systematically overestimate their gas  mass.

\section{CONCLUSIONS}

We  conducted  an  extensive   set   of observations  that     map the
distribution of the  molecular gas in  the inner $\sim 1'\times 1'$ of
the archetypal Seyfert 2/starburst  galaxy NGC 1068. The fully sampled
maps of $ ^{12}$CO, $ ^{13}$CO J=2--1, 3--2  transitions, a $ ^{12}$CO
J=1--0 map and  two test spectra of the  C$ ^{18}$O  J=2--1 transition
allowed us  to  probe the  physical properties  of  the gas   in great
detail. Our main conclusions are the following:

1. A single-phase LVG model  yields average conditions that correspond
   to relatively dense molecular gas ($\sim  10^4$ cm$ ^{-3}$), with a
   temperature of  $\sim 20$~K and  optical depth  $\tau \sim 1-2$ for
   the $ ^{12}$CO J=1--0 transition.  These conditions are not uniform
   and the $ ^{12}$CO $\rm r_{32}=(3-2)/(1-0)$ ratio (at resolution of
   $\sim 16''$) and earlier interferometric  measurements of the  $\rm
   R_{10}=$$\rm ^{12}CO/ ^{13}CO$   and $\rm  R^{(18)} _{10}=$$\rm   C
   ^{18}O/ ^{13}CO$ J=1--0 ratios reveal a wide range of molecular gas
   properties in the inner $\sim 1'\times 1'$  starburst region of NGC
   1068.

2.  A single gas phase with moderate  $ ^{12}$CO J=1--0 optical depths
  cannot reproduce  the  high  C$  ^{18}$O/$ ^{13}$CO  (J=1--0,  2--1)
  intensity ratio and   its   large spatial  variations    observed in
  interferometric     maps  of   the  J=1--0    transition.  A  simple
  two-component model is used to describe the state of the ``average''
  molecular  cloud as consisting of  a dense ($\ga 10^4$ cm$ ^{-3}$),
  more spatially concentrated component where  C$ ^{18}$O J=1--0 has a
  moderate ($\tau \sim 1$) optical depth, surrounded by a more diffuse
  ($\sim 10^3  \rm \  cm ^{-3}$), warmer   gas phase where the $  ^{12}$CO
  J=1--0 transition has $\tau \sim 1-2$.

3. The dense  and more spatially  concentrated  phase contains most of
   the  gas mass.  Since  the $  ^{13}$CO in  this phase  is optically
   thick,  using this isotope to  deduce  molecular gas mass under the
   assumption that it has small/moderate  optical depth throughout the
   CO-emitting volume underestimates the  true mass of this  component
   and hence the total gas mass.

4.  The diffuse  gas phase dominates  the observed $ ^{12}$CO emission,
   it is  highly excited and probably not  virialized.  This leads to
   an overestimate of molecular gas mass when the  luminosity of the $
   ^{12}$CO J=1--0 line and  a standard galactic conversion factor are
   used to  deduce gas mass  in starburst environments.  Since  type 2
   Seyferts harbor starbursts more often than type 1 or field spirals,
   it follows that the use of  the standard galactic conversion factor
   will systematically overestimate  the global molecular gas  mass in
   Seyfert 2's.

We would  like to thank  the  great  crew  of telescope operators  and
support scientists at  the James Clerk  Maxwell Telescope where most of
observations were taken. Special thanks must go to Per Friberg for his
patient  explanations of the telescope's  workings  and to Lorne Avery
and Henry  Matthews  for performing some of   the observations on  our
behalf. We are  greatful  to the  referee Dr   Tamara Helfer  for  her
suggestions that  improved  this  work.  Finally,  we  acknowledge the
support of a research grant to E. R.  S. from the Natural Sciences and
Engineering Research Council of Canada.

This research  has  made use of  the  NASA/IPAC Extragalactic Database
(NED), which is operated by the  Jet Propulsion Laboratory, California
Institute of Technology, under contract  with the National Aeronautics
and Space Administration.

\newpage

\figcaption{The $\rm \langle T_{\rm  mb} \rangle _{\Delta \rm v}$ maps
 of $ ^{12}$CO (top) and $ ^{13}$CO (bottom) of the J=2--1 transition,
 with  $\rm \Delta v  = 950-1300$  km  s$ ^{-1}$,  at a resolution  of
 $\theta  _{\rm i}=24''$.     For  $  ^{12}$CO  the   rms    noise is
 $\sigma_{12}\approx 0.025$~K, the contours are (3,  5, 7,  9, 11,
 13, 15,  17, 19, 21)  x $\sigma  _{12}$, and  the  greyscale range is
 0.075--0.535~K.  For $ ^{13}$CO the rms noise is $\sigma_{13}\approx
 0.003$~K, the contours  are (3, 5,  7, 9,  11, 13, 15)$\times  \sigma
 _{13}$, and the greyscale range is 0.009--0.048~K.
\label{fig1}}

\figcaption{The $\rm \langle T_{\rm mb}  \rangle _{\Delta \rm v}$ maps
 of   $  ^{12}$CO  (top) and $    ^{13}$CO   (bottom) for  the  J=3--2
 transition,   with  $\rm \Delta  v =  950-1300$  km s$   ^{-1}$, at a
 resolution of $\theta  _{\rm i}=16''$.  For  $ ^{12}$CO the rms noise
 is $\sigma_{12}\approx 0.020$~K, the  contours are (3,  5, 7,  9, 11,
 13, 15, 17,  19,  21) x $\sigma   _{12}$, and the greyscale range  is
 0.060--0.456~K.  For $  ^{13}$CO the rms noise is $\sigma_{13}\approx
 0.002$ K, the contours are (3, 5, 7, 9, 11, 13)$\times \sigma _{13}$,
 and the greyscale range is 0.006--0.026~K.
\label{fig2}}

\figcaption{Channel maps of  the $ ^{12}$CO J=3--2 emission (contours)
 overlayed on the $ ^{12}$CO  J=1--0 emission (greyscale) at a  common
 spatial resolution of 16$''$ and velocity resolution of $\rm \Delta v
 _{\rm  chan}\sim  8$ km  s$  ^{-1}$. The  average  rms noise of the $
 ^{12}$CO J=3--2 map is $\sigma _{\rm T}(3-2)=0.065$ K, the lowest level
 contour is  3 $\sigma _{\rm T}(3-2)$  and  the contour  interval is 2
 $\sigma _{\rm T}(3-2)$.   The rms noise  of the $ ^{12}$CO J=1--0 map
 is $\sigma _{\rm T}(1-0)=0.08$ K and the greyscale range is: 0.25--2.50
 K.  The line ratio is estimated as $\rm r_{32} ^{(\rm chan)}=\langle
 T_{\rm mb}(3-2)  \rangle _{\rm R_L}/\langle  T_{\rm mb}(1-0)  \rangle
 _{\rm R_L}$ for  a radius  of  $\rm R_{\rm L}\la  30''$  from the map
 center and it is  reported at the bottom  left corner of each channel
 map.  The beam is  shown  at the bottom   right  corner of the  first
 channel map.
\label{fig3}}

\figcaption{Spectra of C$ ^{18}$O, $ ^{13}$CO J=2--1 in two positions
The offsets from the map center are the following
 A: $(\Delta a, \Delta \delta)=(8'', 0'')$, 
 B: $(\Delta a, \Delta \delta)=(-8'', -3'')$
\label{fig4}}

\figcaption{
Area-averaged spectra of $ ^{12}$CO, $ ^{13}$CO for J=3--2, 2--1, 1--0,
over a circular area with radius $\rm R=30''$.
\label{fig5}}

\newpage
\centerline{\large Table 1}
\centerline{\large NGC 1068: The observations}
\begin{center}
\begin{tabular}{ c c c c c c c } \hline\hline

Run & Receiver & Spectral line  &  $\Delta \rm v$$ ^{\rm c}$  &
 $\theta _{\rm HPBW}$$ ^{\rm d}$  & $\Delta \theta _{\rm s}$$ ^{\rm e}$ &
  Map size  \\ 
  
 & ($\rm T_{\rm sys}$)$ ^{\rm a}$  & ($\nu _{\circ}$)$ ^{\rm b}$  & ($\Delta \rm v_{\rm chan}$)$ ^{\rm c}$  &
($\delta \theta _{\rm rms}$)$ ^{\rm d}$ & ($\rm T_{\rm int}$)$ ^{\rm e}$  & (\# of points)$ ^{\rm f}$ \\ \hline\hline

1994, Jan. 2-8 & A2  & $ ^{12}$CO J=2--1     &  900   & $21''$  & $10''$ & $100''\times 60''$\\ 
1994, Nov. 26-29    & (350-450) & (230.538) & (0.81) & ($4''$) & (1-2)  & (85) \\ \hline

1994, Jan. 2-8 & A2  & $ ^{13}$CO J=2--1  & 941  & $21''$ & $10''$ & $80''\times 60''$\\
1994, Nov. 26-29    &  (400-500)         & (220.398)    & (0.85) & ($4''$) & (10-20) & (57)\\ \hline

1997, Nov. 27 & A2  & C$ ^{18}$O J=2--1  & 945      & $22''$ & ... & ...\\
                        & (500)    & (219.560)       & (0.85)   & ($4''$)  & (90, 110) & (2)\\ \hline

1994, Nov. 26-29   & B3i & $ ^{12}$CO J=3--2 & 600 & $14''$ & $7''$   & $80''\times 60''$\\
1996, Jan. 16-28   & (750-1000) & (345.795)  & (0.54)       & ($3''$) & (2) & (113)\\ \hline

1997, Nov. 27-Dec. 1  & B3 & $ ^{13}$CO J=3--2 & 900 & $14''$ & $7''$  & $60''\times 50''$\\
1997, Dec. 27-29      & (400-500) & (330.588) & (0.57) & ($3''$) & (10-15) & (87)\\ \hline\hline

\end{tabular}
\end{center}

$ ^{\rm a}$ Typical system temperatures for  the corresponding receiver in Kelvin.

$ ^{\rm b}$ The rest frequency of the spectral line in GHz.

$ ^{\rm c}$ The corresponding velocity range $\Delta \rm v$, and resolution $\Delta \rm v_{\rm chan}$
   in km s$ ^{-1}$.

$ ^{\rm d}$ The HPBW of the gaussian beam,  and the rms pointing error $\delta \theta _{\rm rms}$.

$ ^{\rm e}$ The sampling interval $\Delta \theta _{\rm s}$, and the integration time
$\rm T_{\rm int}$ per point in minutes.

$ ^{\rm f}$ The number of points corresponding to each map.

\newpage 
\centerline{\large Table 2}
\centerline{\large NGC 1068: The global line ratios}
\begin{center}
\begin{tabular}{| c | c | c | c | c | c | c |} \hline

$\frac{\rm ^{12}CO(3-2)}{\rm ^{12}CO(1-0)}$ & $\frac{\rm ^{12}CO(2-1)}{\rm ^{12}CO(1-0)}$ &
$\frac {\rm ^{12}CO(3-2)}{\rm ^{13}CO(3-2)}$ & $\frac {\rm ^{12}CO(2-1)}{\rm ^{13}CO(2-1)}$
& $\frac {\rm ^{12}CO(1-0)}{\rm ^{13}CO(1-0)}$ & $\frac {\rm C ^{18}O(1-0)}{\rm ^{13}CO(1-0)}$ & $\frac {\rm C ^{18}O(2-1)}{\rm ^{13}CO(2-1)}$\\ 

($\rm r_{32}$) & ($\rm r_{21}$) & ($\rm R_{32}$) & ($\rm R_{21}$) & ($\rm R_{10}$) & 
($\rm R^{(18)} _{10}$) & ($\rm R^{(18)} _{21}$)\\ \hline

$0.52\pm 0.17$ & $0.68\pm 0.23$ & $14\pm 3$ & $10\pm 2$ & $14\pm 2\ ^{\rm a}$ 
& $0.30\pm 0.10\ ^{\rm b}$ & $0.23\pm 0.04\ ^{\rm c}$\\ \hline
\end{tabular}
\end{center}
 $ ^{\rm a}$ OVRO measurement (Papadopoulos, Seaquist \& Scoville 1996),  Young \& Sanders (1986).\newline
\vspace*{0.5cm}
\hspace*{-0.35cm} $ ^{\rm b}$ OVRO measurement (Papadopoulos, Seaquist \& Scoville 1996).\newline
\vspace*{-1.6cm}

\noindent
 $ ^{\rm c}$ The average of the two values measured with the JCMT.

\newpage

\vspace*{-0.5cm}
\centerline{\large Table 3}
\centerline{\large NGC 1068: Two-phase model}

\begin{center}
\begin{tabular}{| c | c | c | c |} \hline\hline
Gas phase  (B)$ ^{\rm a}$  & Line ratios (A)$ ^{\rm b}$   & LVG parameters  (A)$
 ^{\rm c}$ & Excitation
(A)$ ^{\rm d}$\\ $\rm \tau   ^{(18)} _{10}$, $\rm  f_c  \rho _{13}$, $\rm  T_{\rm
kin}$ &  $\rm r_{21}$, $\rm  r_{32}$,  $\rm R_{10}, R_{21},  R_{32}$  &
$\rm  T_{\rm   kin}$, $\rm n(H_2)$,  $\Lambda$ &  $\rm
E^{(13)} _{10}$, $\tau^{(12)} _{10}$\\ \hline\hline

0.5, 1.35, 15     & 0.97, 0.74, 32, 15, 31   & 35, $\rm 3 x 10^3$, $\rm 1 x 10^{
-6}$ & 3.8, 0.29 \\
                  & (1.18, 0.73, 27, 23, 31) &      $\chi^2  =3.2$ &    \\ \hline
0.7, 0.70, 10     & 0.97, 0.74, 23, 13, 20   & 35, $\rm 3 x 10^3$, $\rm 3 x 10^{-
6}$ & 2.8, 0.94\\
                  & (1.05, 0.76, 17, 13, 20) &      $\chi^2  =3.2$ & \\ \hline
0.9, 0.50, 15     & 0.97, 0.74, 20, 12, 20   & 35, $\rm 3 x 10^3$, $\rm 3 x 10^{-
6}$ & 2.8, 0.94\\
                  & (1.05, 0.76, 17, 13, 20) &     $\chi^2  =0.95$ & \\ \hline
1.1, 0.40, 20     & 0.97, 0.74, 19, 12, 20   & 35, $\rm 3 x 10^3$, $\rm 3 x 10^{-
6}$ & 2.8, 0.94\\
                  & (1.05, 0.76, 17, 13, 20) &     $\chi^2  =0.53$ & \\ \hline
1.5, 0.30, 20-40  & 0.97, 0.76, 18, 12, 19   & 35, $\rm 3 x 10^3$, $\rm 3 x 10^{-
6}$ & 2.8, 0.94\\
                  & (1.05, 0.76, 17, 13, 20) &     $\chi^2  = 0.37-0.45$ & \\ \hline
2.5, 0.25, 15-40  & 0.97, 0.74, 17, 11, 17-18 &  35, $\rm 3 x 10^3$, $\rm 3 x 10^
{-6}$ & 2.8, 0.94\\
                  & (1.05, 0.76, 17, 13, 20)  & $\chi^2 =1.2-0.8$ & \\ \hline
\end{tabular}
\end{center}

\begin{small}
\noindent
$ ^{\rm a}$ The physical  conditions of the  gas  in phase (B):  $\tau
^{(18)} _{10}$  is  the optical    depth of  the   C$ ^{18}$O   J=1--0
\hspace*{0.2cm} transition, $\rm T_{\rm kin}$ is  the kinetic temperature and $\rm f_c
\rho _{13}$ is defined in Equation (3).

\noindent
$ ^{\rm  b}$ The line ratios estimated  for gas  phase (A),
 the  ones in the parenthesis  are the values obtained
\hspace*{0.2cm} from  the LVG model with the  minimum $\chi ^2$, where
$\chi ^2 = \sum _{\rm i} \frac{1}{\sigma _{\rm i} ^2}\left[\rm R_{(\rm
i)}-R_{\rm obs}\right]^2$ and $\rm R_{(\rm i)}$, $\rm R_{\rm obs}$ are
\hspace*{0.2cm} the model  and observed line ratios respectively, with
$\sigma _{\rm i}$ being the corresponding 1$\sigma $ uncertainty.

\noindent
$ ^{\rm c}$ The  LVG parameters (see  text) corresponding to  the best
 fit of the line ratios of gas phase (A).

\noindent
$ ^{\rm  d}$ $\rm  E^{(13)} _{10}=T_{\rm exc}/T_{\rm  kin}$  for  $
^{13}$CO J=1--0  and $\tau ^{(12)}  _{10}$ is the optical depth  of
the $^{12}$CO  J=1--0  transition \hspace*{0.2cm} for gas phase (A).

\end{small}

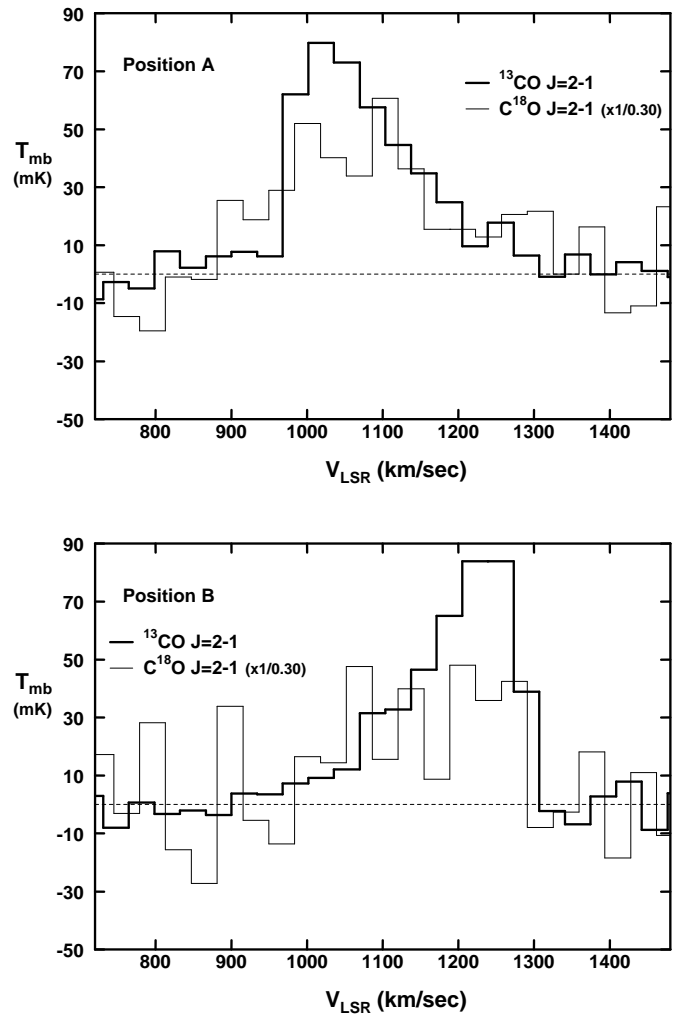
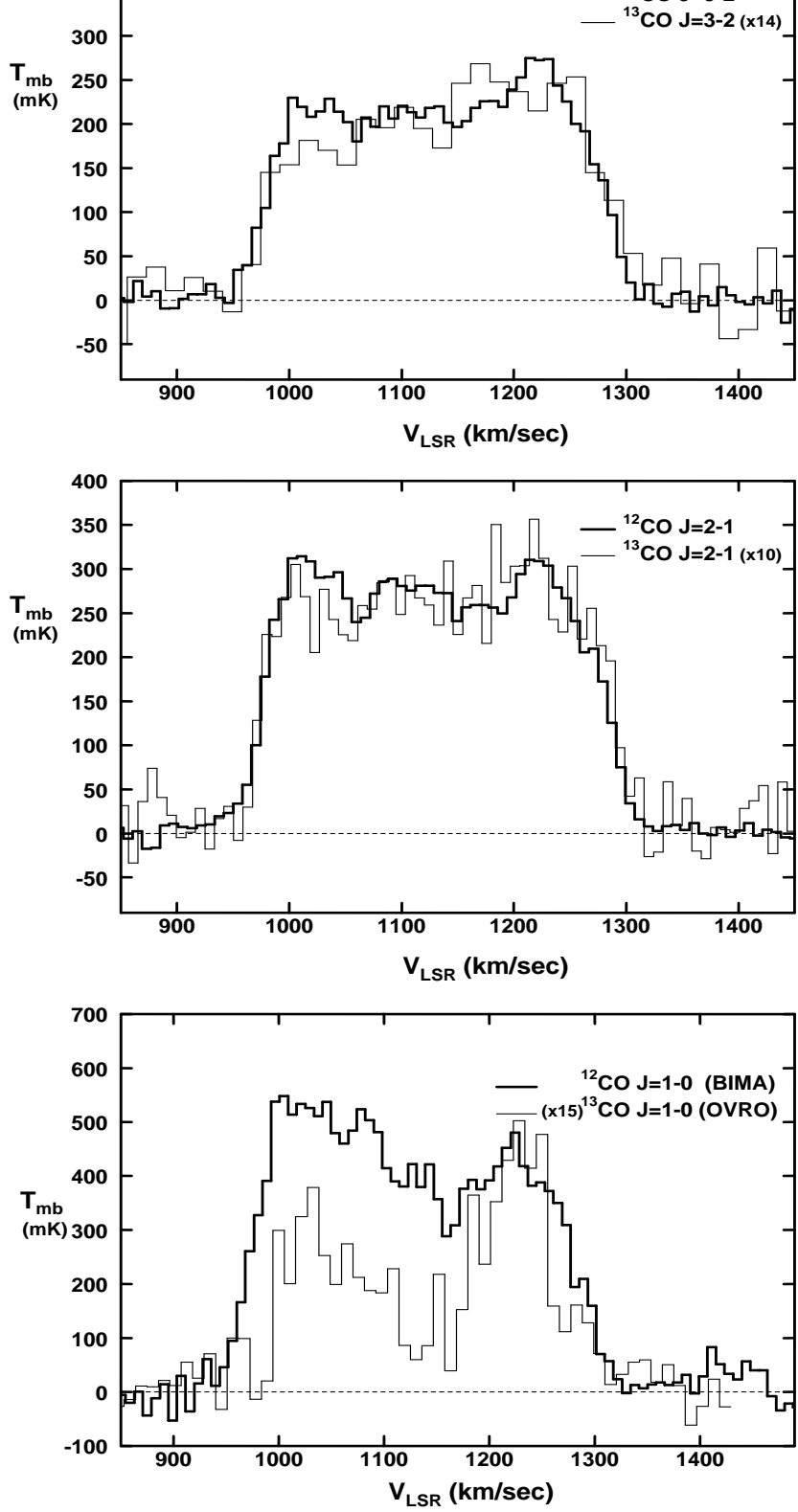
\end{document}